\documentclass{article}

   \usepackage[final]{neurips_2019}

\usepackage{verbatim}
\usepackage[utf8]{inputenc} 
\usepackage[T1]{fontenc}    
\usepackage{hyperref}       
\usepackage{url}            
\usepackage{booktabs}       
\usepackage{amsfonts}       
\usepackage{nicefrac}       
\usepackage{microtype}      
\usepackage[pdftex]{graphicx}
\title{Unsupervised Sparse-view Backprojection via Convolutional and Spatial Transformer Networks}

\author{%
  Xueqing Liu, Paul Sajda\\
  Department of Biomedical Engineering\\
  Columbia University\\
  New York, NY 10027 \\
  \texttt{xl2556@columbia.edu, psajda@columbia.edu} \\
}

\begin{document}

\maketitle

\begin{abstract}
  Many imaging technologies rely on tomographic reconstruction,  which  requires solving a multidimensional inverse problem given a finite number of projections. Backprojection is a popular class of algorithm for tomographic reconstruction, however it typically results in poor image reconstructions when the projection angles are sparse and/or if the sensors characteristics are not uniform.  Several deep learning based algorithms have been developed to solve this inverse problem and reconstruct the image using a limited number of projections. However these algorithms typically require examples of the ground-truth (i.e. examples of reconstructed images) to yield  good performance. In this paper, we introduce an unsupervised sparse-view backprojection algorithm, which does not require ground-truth. The algorithm consists of two modules in a generator-projector framework; a convolutional neural network and a spatial transformer network. We evaluated our algorithm using computed tomography (CT) images of the human chest. We show that our algorithm significantly out-performs filtered backprojection when the projection angles are very sparse, as well as when the sensor characteristics vary for different angles.  Our approach has practical applications for medical imaging and other imaging modalities (e.g. radar) where sparse and/or non-uniform projections may be acquired due to time or sampling constraints.       
\end{abstract}

\section{Introduction}
Medical imaging modalities, such as computed tomography (CT), are based on acquiring a finite numbers of projections given detectors (e.g. X-ray detector) and generators/sources (e.g. X-ray source). The projections are usually acquired by rotating the source (and/or detector) around the object of interest and then acquiring projections at different angles. To reconstruct the structure of the object being observed, the multidimensional inverse problem  must be solved. Two conventional, and widely used reconstruction methods are backprojection and the Fourier-domain reconstruction algorithm. In fact these two methods are tightly related and it has been shown for example that applying a filter in the Fourier-domain will yield similar results to filtered backprojection[\citenum{brooks1975theory}]. We thus refer to these two classes of reconstruction methods more generally as "backprojection algorithms".  Another class of  reconstruction method is iterative reconstruction[\citenum{beister2012iterative}]. Iterative reconstruction methods make strong assumptions about the scanner geometry, scanner optics and noise statistics, and implement these assumptions as constraints within a multiple iteration optimization procedure.

Convolution neural networks (CNN) have been applied to multiple image processing tasks such as denoising[\citenum{zhang2017beyond}], super-resolution[\citenum{kim2016accurate}], etc. CNNs have been used in the context of image reconstruction via backprojection, though mostly as a form of image post-processing to improve SNR and image quality[\citenum{kang2018deep}][\citenum{jin2017deep}][\citenum{han2018framing}][\citenum{chen2017low}].  CNNs have also been used in iterative reconstruction loops[\citenum{ye2019deep}]. One innovative way that CNNs have been used is in cases of sparse sensors/projections.  For example in [\citenum{ye2018deep}], the authors describe an end-to-end deep learning model to reconstruct CT images given a limited number of projections, with results indicating better reconstructions than conventional backprojection. However their method relies on a training set that includes  high resolution reconstructed images as the ground-truth. The ground-truth thus requires a high density of projections and one of the conventional methods for reconstructing images from these projections.

In this paper, we propose an unsupervised backprojection algorithm that combines a CNN and a Spatial Transformer Network (STN) [\citenum{jaderberg2015spatial}] inspired network to enable both sparse-view reconstruction as well as reconstruction when sensors (e.g. X-ray detectors and/or sources) are non-uniform for different angles. The CNN acts as the generator and the STN inspired network as the projector in a generator-projector architecture. The original STNs introduced in [\citenum{jaderberg2015spatial}] are able to learn affine transformation parameters and implement differentiable geometric transformation and are thus well-suited to be modified to a projector. Novel to our generator-projector architecture is that it can be trained without the need for ground-truth from high resolution reconstructions--i.e. it is sparse-view both in terms of training and testing. We conduct evaluations of the algorithm using chest CT data and show the algorithm outperforms conventional backprojection algorithms, specifically in cases having very sparse projections. We also show that when  sensors for obtaining the projections are of non-uniform characteristics, the improvement beyond conventional backprojection algorithms is considerable.

\section{Unsupervised Deep Backprojection} \label{Unsupervised Deep backprojection}

\begin{figure}\label{framework}
  \centering
  \includegraphics[scale=0.275]{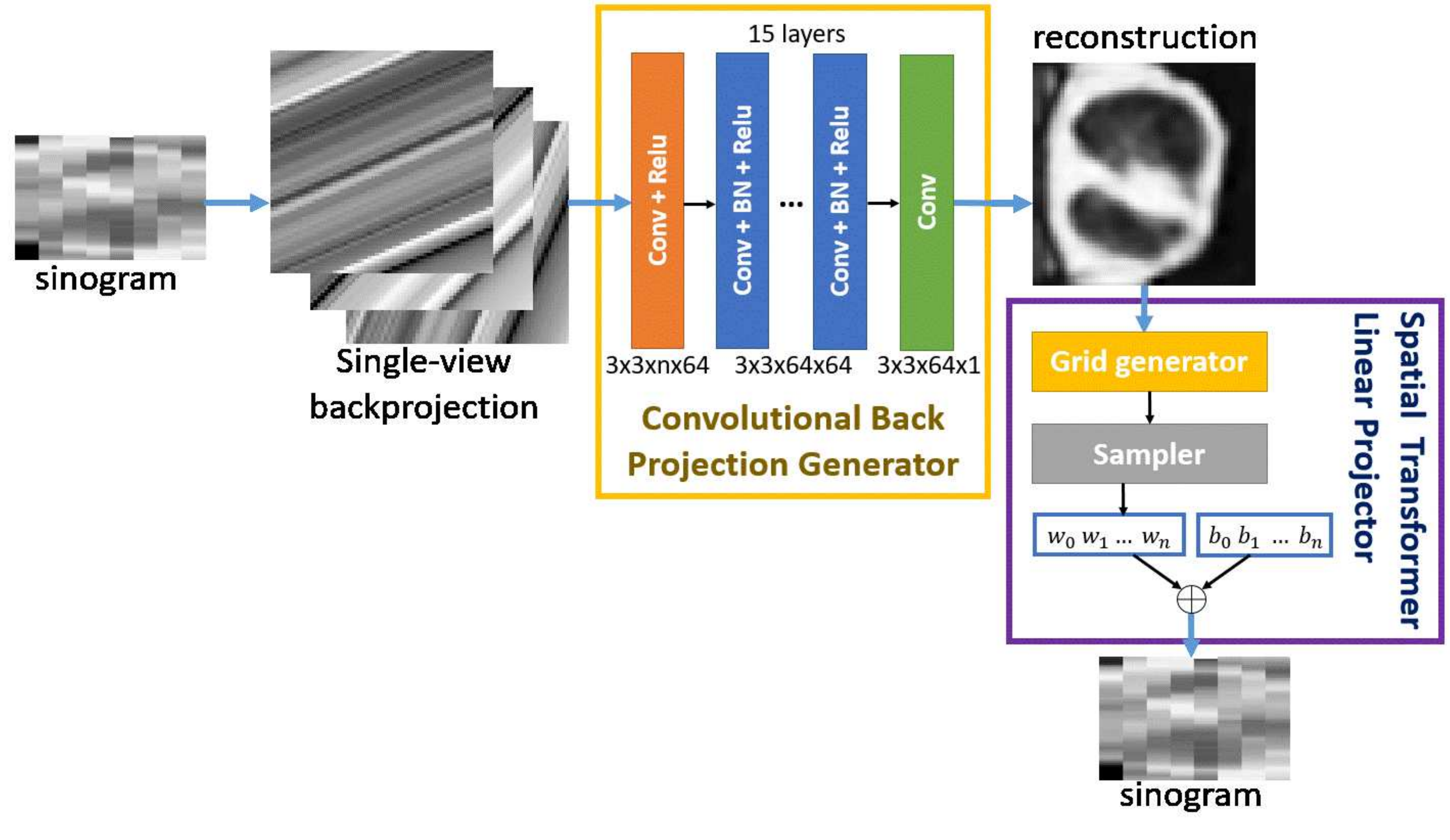}
  \caption{Framework of unsupervised sparse-view backprojection: The input and output of the network are both sparse-view sinograms. A sinogram is converted to single-view backprojections which serve as input to the convolutional backprojection generator. The generator outputs the backprojection reconstruction. The spatial transformer linear projector takes the reconstruction and transformation angle as inputs and generates a sinogram. The projector also adjusts to sensor non-uniformity with trainable weights ($w_i$) and biases ($b_i$) }
\end{figure}

 Fig.~\ref{framework}, is an overview of our framework.  Sinograms are converted to single-view backprojections before feeding into a multi-layer CNN generator. The generator predicts the reconstruction from the sparsely measured sinograms, while the projector maps the reconstruction back to the original sinograms. 
 
\begin{figure}\label{singleview}
  \centering
  \includegraphics[scale=0.14]{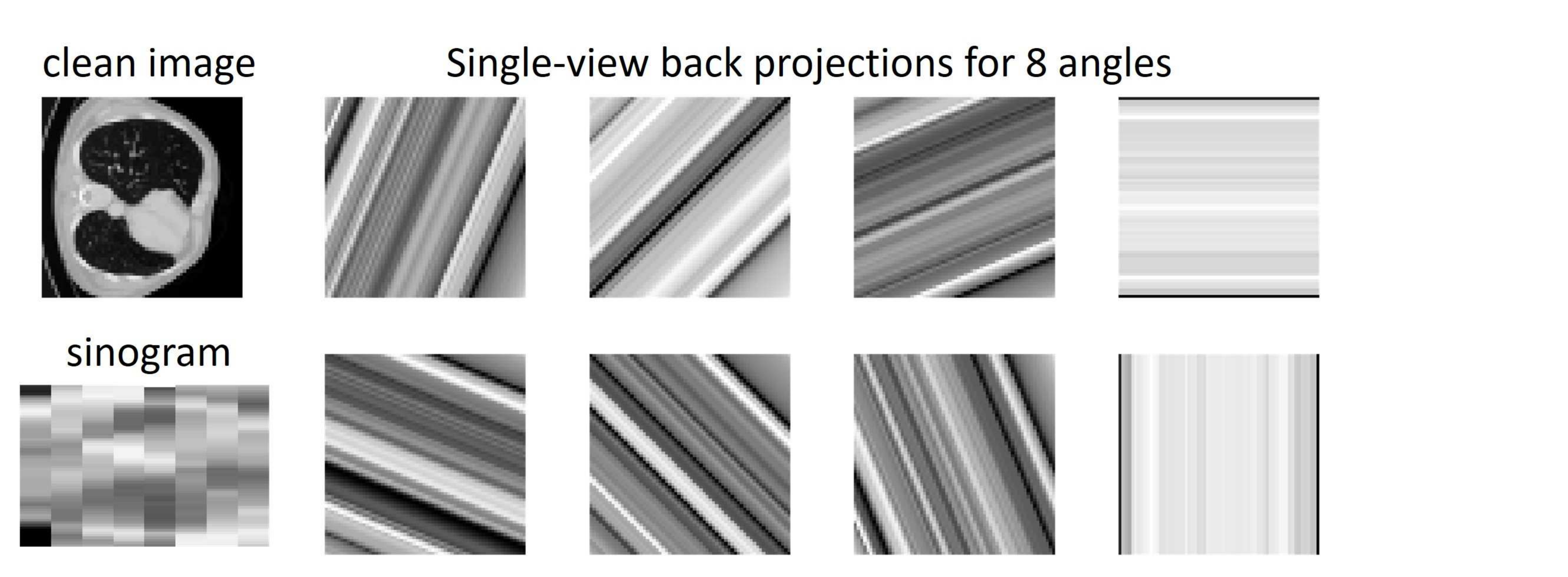}
  \caption{Example of a high resolution CT image (upper left), its 8-angle projection sinogram (lower left) and its single-view backprojections (right). Note that the high resolution CT image is constructed from hundreds of projections, and would normally serve as ground-truth in other deep learning based backprojection algorithms.}
\end{figure}

\subsection{Single-view Backprojections}

The relationship between the data space $S$ and its set of projections is defined as the  Radon transformation:

\begin{equation}\label{Radon}
Radon(S, \theta) = R_S(l(\theta), \theta) = \int_{-\infty}^{\infty}\int_{-\infty}^{\infty}S(x,y)\delta(l(\theta)-xcos\theta-ysin\theta)dxdy
\end{equation}

where $\delta(\cdot)$ is the Dirac delta function and $l(\theta)=xcos\theta+ysin\theta$.

Specifically, one projection $p_i$ as an integration of the data space $S$ along a particular direction $\theta_i$, can be represented with Radon transform as
\begin{equation}\label{one}
p_i=Radon(S, \theta_i)
\end{equation}
A sinogram, a standard data structure to store projections, is defined as the $m\times n$ matrix made of $n$ projections $(p_1, p_2, p_3, ..., p_n)$ from different angles$(\theta_1, \theta_2, \theta_3, ..., \theta_n)$.  Backprojection tries to solve the inverse problem, namely to reconstruct the data $S$ from the sinograms generated from its limited numbers of projections $(p_1, p_2, p_3, ..., p_n)$.

Instead of using sinograms directly as input to a CNN we construct single-view backprojections[\citenum{ye2018deep}]. This is done by performing single projection backprojection and stacking the backprojection results. An example of 8 angle single-view backprojections is shown in Figure \ref{singleview}. 

\subsection{Convolutional Backprojection Generator}

The stacked $n$ (where $n$ is the number of projections) single-view backprojections serve as the $n$-channel input to the convolutional backprojection generator. 

The convolutional backprojection generator is composed of 17 convolution layers. Batch normalization is applied to all layers except for the first and last layers. ReLU activation functions are applied to all layers except the last layer. Each of the layers except the last uses 64 convolution kernels of size $3\times 3$. The last layer has 1 kernel of size $3\times 3$ to construct the backprojection prediction from all inputs.

\subsection{Spatial Transformer Linear Projector}
The backprojection reconstruction serves as input to an STN inspired linear projector to generate predicted sinograms. The projector applies the Radon transform of correspondent angles $(\theta_1, \theta_2, \theta_3, ..., \theta_n)$ to the backprojection reconstruction to regenerate the sinogram prediction as in equation (\ref{Radonhat}). To implement a  differentiable Radon transformation that allows gradient-based back propagation, we use the spatial transformers introduced in [\citenum{jaderberg2015spatial}]. The grid generator transforms the original regular spatial grid of the reconstruction to a sampling grid. The sampler produces the sampled transformed data from the reconstruction at the grid points. Then a trainable linear mapping, as in equation (\ref{linear}), is applied to each $\hat{p_i}$ with different $w_i$ and $b_i$, which compensates for possible sensor non-uniformity.

\begin{equation}\label{Radonhat}
\hat{p_i} = Radon(\hat{S}, \theta_i)
\end{equation}

\begin{equation}\label{linear}
\hat{{p_i}'} = w_i \hat{p_i}+b_i
\end{equation}

The objective function is given in equation (\ref{obj}). We minimize the mean squared error between the generated sinogram $\hat{{P}'}= (\hat{{p_1}'},...,\hat{{p_n}'})$ and the sinogram ground-truth $P=(p_1,...,p_n)$. Note that this ground-truth is different from the ground-truth used in  [\citenum{ye2018deep}] since in our case the ground-truth is the sparse projections whereas in [\citenum{ye2018deep}]  it is the high-resolution (i.e. large number of) projections and the corresponding reconstructed image.  We also include an l1-norm of predicted backprojection reconstruction $\hat{S}$ to impose a sparse reconstruction.  

\begin{equation}\label{obj}
l(k, w, b) = \frac{1}{n}\sum^{n}_{k=1}||p_i - \hat{{p_i}'}||_2^2+\alpha ||\hat{S}||_1
\end{equation}

\begin{figure}\label{uniform}
  \centering
  \includegraphics[scale=0.37]{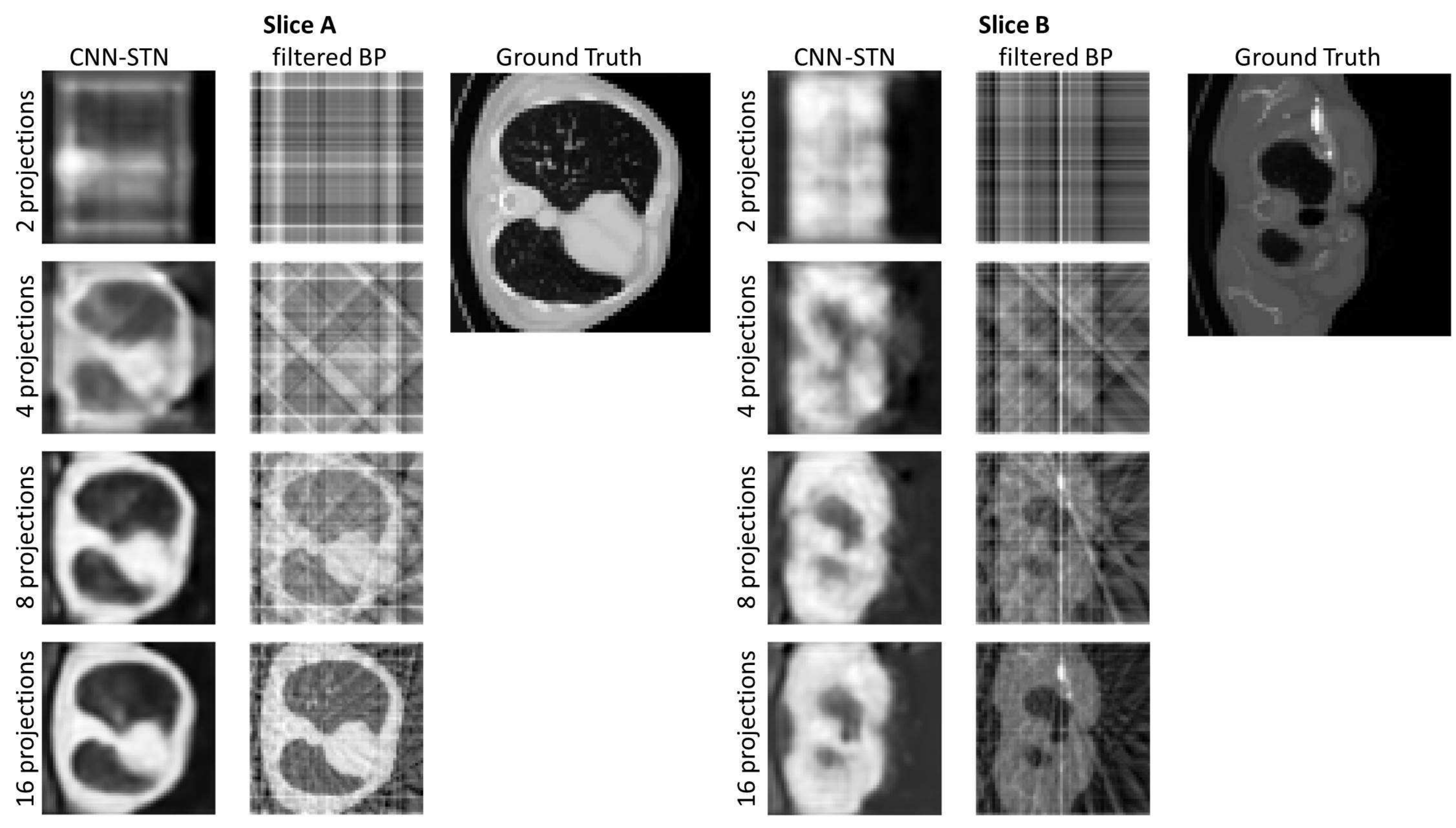}
  \caption{ Reconstruction results for the case of uniform sensors: A and  B are two example slices of a reconstructed chest CT scan. For each example slice, the left most column is the reconstruction result using our algorithm, the middle column is the reconstruction result using filtered backprojection and the right image is the ground truth. }
\end{figure}

\section{Experiment and Results}
We evaluate our algorithm using 43 human chest CT scans [\citenum{Dataset}]. The dataset is part of The Cancer Imaging Archive (TCIA)[\citenum{clark2013cancer}]. 2/4/8/16-angle sinograms are generated by applying the corresponding Radon transformation to each slice of the CT data with and without sensor non-uniformity respectively. Sensor non-uniformity is introduced by multiplying each projection $p_i$ with a weight $w_i$ and adding a bias $b_i$. Both $w_i$ and $b_i$ are random numbers with a standard normal distribution. $w_i$ and $b_i$ are constants for different slices of the same scan.   

\begin{figure}\label{nonuniform}
  \centering
  \includegraphics[scale=0.37]{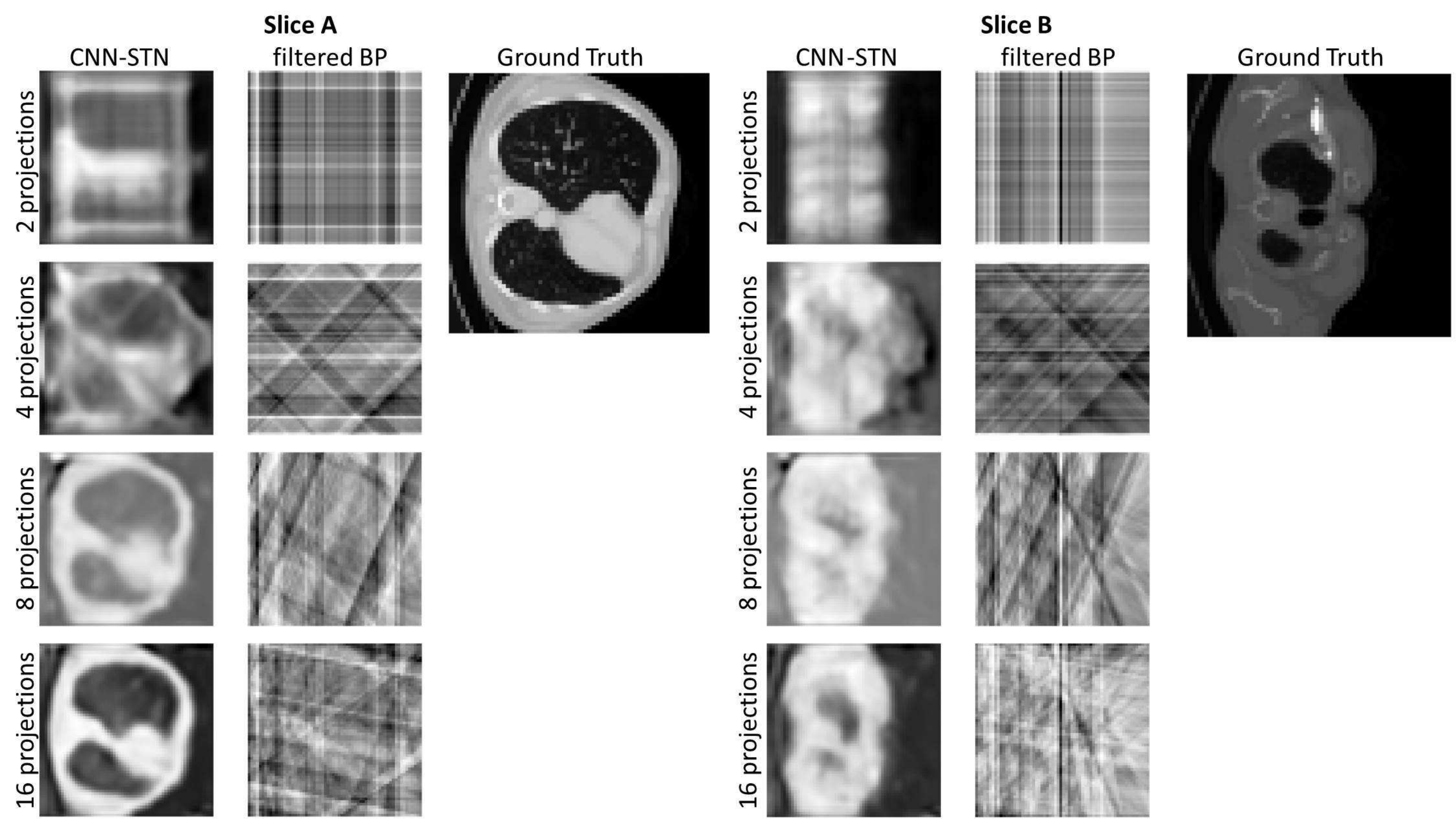}
  \caption{Reconstruction results for the case of non-uniform sensors: A and  B are two example slices of a reconstructed chest CT scan. For each example slice, the left most column is the reconstruction result using our algorithm, the middle column is the reconstruction result using filtered backprojection and the right image is the ground-truth.}
\end{figure}

To test the performance of our unsupervised algorithm on very limited data, each model is trained and tested on the same subject's scan. The mean number of slices for each scan is $80.70\pm 21.16$. We compare the performance of our algorithm and filtered backprojection on 2/4/8/16-angle sinograms reconstruction with and without sensor non-uniformity. When testing without sensor non-uniformity, we fix the weights $w_i=1$ and bias $b_i=0$ for the STN inspired projector. 

Figure 3 shows results for two different slices given 2/4/8/16-angle projection reconstructions.  We compare  our algorithm to filtered backprojection assuming sensor uniformity (i.e. fixed $w_i=1$ and $b_i=0$). In all cases, our algorithm performs better than filtered backprojection. The improved performance is especially apparent in very sparse cases - 2/4-angle projection reconstructions of filtered backprojection barely show any useful information while our algorithm can still provide meaningful results. To compare the performance in an objective way, we calculated the mean square error(MSE) of the reconstruction as in  equation (\ref{MSE}), where $S$ stands for the ground truth image and $\hat{S}$ is the reconstruction. We then calculated reconstruction peak signal-to-noise ratio (PSNR) on top of MSE(shown in equation (\ref{PSNR})). Here $MAX_S$ stands for the maximum posible pixel value of ground truth $S$. 

\begin{equation}\label{MSE}
MSE = \frac{1}{mn}\sum_{i=0}^{m-1}\sum_{j=0}^{n-1}[S(i,j)-\hat{S}(i,j)]^2
\end{equation}

\begin{equation}\label{PSNR}
PSNR = 20 log_{10}(\frac{MAX_S}{\sqrt{MSE}})
\end{equation}

A comparison of reconstruction PSNR between our algorithm and filtered backprojection is shown in Figure 5.a. For 2/4/8/16 angles of projections, our algorithm shows significantly better PSNR than filtered backprojection.

Figure 4 shows two slices' 2/4/8/16-angle projection reconstruction results for our algorithm compared with filtered backprojection for the case of  sensor non-uniformity. Filtered backprojection cannot adjust to sensor non-uniformity and shows strong artifacts. Our algorithm can suppress the impact of sensor non-uniformity and provides reasonable reconstructions. Due to sensor non-uniformity, the intensity and contrast of the original image cannot be recovered, hence a PSNR comparison is meaningless. Instead, we use the correlation coefficient between ground-truth and reconstructed images as an evaluation metric of the reconstruction performance. Figure 5.b shows the correlation coefficients of our algorithm and filtered backprojection. Our algorithm significantly outperforms filtered  backprojection, with performance continuing to  improves as  the number of projections acquired increases.  Note that this is not true for filtered backprojection, where performance  does not improve with more projections when sensors are  non-uniform. 

\begin{figure}\label{result}
  \centering
  \includegraphics[scale=0.64]{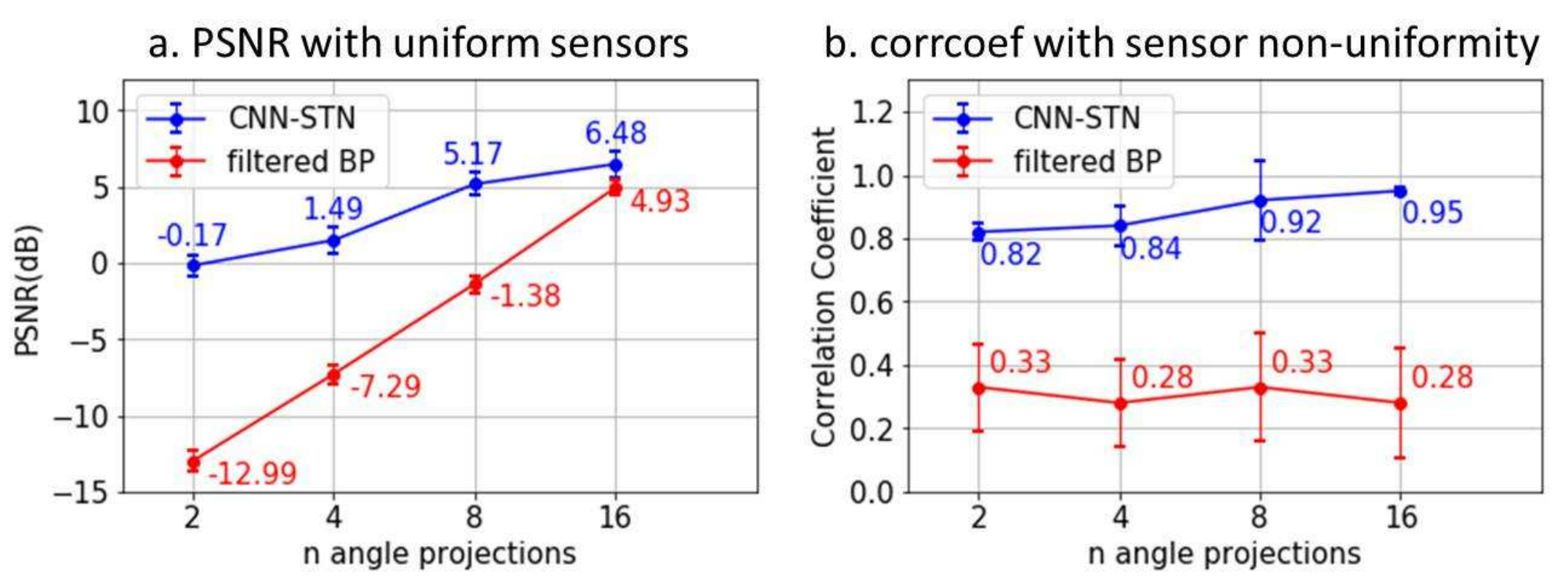}
  \caption{Comparison of our algorithm with filtered backprojection: (a) shows the reconstruction PSNR of our algorithm and filtered backprojection with uniform sensors. (b) shows the  reconstruction correlation coefficient of our algorithm and filtered backprojection non-uniform sensors.}
\end{figure}

\section{Discussion} \label{discussion}
In this paper, we introduced an unsupervised backprojection algorithm using a generator-projector framework based on a CNN and STN. Our results show much better performance than the conventional filtered backprojection algorithm. In general we show that deep learning models, with relatively few parameters as we have in the CNN and STN, can be applied to unsupervised tasks that have very limited training data. 

It is to be noted as shown in Figure 5.a, with uniform sensors, when the number of projections increases, the performance of our algorithm increases more slowly than filtered backprojection. With dense projections, filtered backprojection still provides more accurate reconstruction results. However our approach enables methods which are limited in the number of projections or where sensor characteristic is arbitrary or non-uniform.   Finally, this algorithm can be applied more broadly, for example to  applications when the sensors are sparse and non-uniform, and/or there is no ground-truth, as would be required in supervised learning. For example, radar and visual input based reconstruction have limited numbers of sensors/projections. Acquiring the ground-truth for training is also expensive and can require human effort/labeling. Sensor non-uniformity is common for these applications because multiple sensors are often used for different angles. Another potential application comes from [\citenum{transcoding}] where the author investigated using CNNs to transcode EEG to fMRI and vice versa using simultaneously acquired EEG-fMRI data -- i.e. predicting one modality from another. The overall hypothesis is that a joint transcoding/prediction will enable the reconvery of the latent source space. According to the author, one of the difficulties of the problem is the absence of ground-truth for the source space. After investigating the problem, we figure that the EEG predicted source space and fMRI predicted source space can be considered as two projections of the same source space and since the two modalities will yield different intensity scales of the source space predictions, we  can consider it a backprojection problem with sensor non-uniformity.  Application of our algorithm to this neural source reconstruction problem is a research direction we are currently pursuing.

\section{Acknowledgements}
This work is supported by the Army Research Laboratory under Cooperative
agreement number W911NF-10-2-0022

\bibliographystyle{IEEEtranN}
\bibliography{main.bib} 

\end{document}